\documentclass[doublecol,linenumbers]{epl2} 
\usepackage{amsmath}

\usepackage[english]{babel}
\usepackage{graphicx}
\usepackage{epsfig}
\usepackage{amssymb}
\usepackage{dcolumn}
\usepackage{bm}
\usepackage{color}
\usepackage{float}
\usepackage{hyperref} 
\hypersetup{colorlinks=true}

\title{Influence of  Lifshitz transitions and correlation effects on the scattering rates of the charge carriers in iron-based superconductors.}
\shorttitle{Lifshitz transitions and correlation effects}

\author{J. Fink\inst{1}}
\shortauthor{J. Fink}

\institute{                    
  \inst{1}Leibniz Institute for Solid State and Materials Research  Dresden, Helmholtzstr. 20, D-01069 Dresden, Germany\\
}

\newcommand{\BFCA}{\ensuremath{\mathrm{Ba(Fe}_{1-x}\mathrm{Co}_{x})_{2}\mathrm{As}_{2}}}

\newcommand{\BFAP}{\ensuremath{\mathrm{BaFe}_{2}\mathrm{As}_{2-x}\mathrm{P}_{x}}}

\abstract{
Minimum model calculations on the co-action of hole vanishing Lifshitz transitions and correlation effects in ferropnictides are presented. The calculations predict non-Fermi-liquid behaviour and huge mass enhancements of the charge carriers at the Fermi level. The findings are compared with recent ARPES experiments and with measurements of  transport  and thermal properties of ferropnictides. The results from the calculation can be also applied to other unconventional superconductors and question the traditional view of quantum critical points.}

\begin{document}

\maketitle

\section{Introduction }

Bad metals  and unconventional superconductivity are both one of the most active research fields in solid state physics. Extensive research efforts have revealed that these phenomena appear near particular points, termed quantum critical points (QCP),  in the phase diagram temperature vs. a control parameter. These control parameters can  be pressure, chemical pressure, doping, or magnetic field. Near the QCP, the normal state properties are characterized by non-Fermi liquid behaviour, dubbed bad metal. At low temperatures very often unconventional superconductivity is observed. Since the QCP  appears at the end of an antiferromagnetic range, a widespread view of these phenomena is related to a scenario in which  at low temperatures the charge carriers are strongly coupled to antiferromagnetic quantum fluctuations\,\cite{Loehneysen2007,Gegenwart2008,Haslinger2002}. These fluctuations could be the glue mediating unconventional superconductivity. They  could also account for the strange normal state non-Fermi-liquid behaviour as visible in transport  and thermal properties of ferropnictides.
The former have manifested themselves e.g. by a linear temperature dependence of the resistivity which was observed in various highly correlated systems such as heavy fermion systems\,\cite{Custers2003}, cuprates\,\cite{Daou2009}, and iron based superconductors\,\cite{Kasahara2010a,Analytis2014}. In this context we mention the phenomenological model of a marginal Fermi-liquid, sometimes also called singular liquids, which was developed to understand Raman experiments and transport properties of doped cuprates\,\cite{Varma2002}.

Besides transport and thermal  properties,  angle-resolved photoemission spectroscopy (ARPES)\,\cite{Damascelli2003}  is a versatile method to study the electronic structure of solids. In the unconventional superconducting ferropnictides \,\cite{Johnston2010} near optimal doping, ARPES has detected changes of the topology of Fermi surfaces (Lifshitz transitions\,\cite{Lifshitz1960} of the pocket vanishing type) for the electron doped systems \,\cite{Malaeb2009,He2010,Thirupathaiah2010,Thirupathaiah2011,Liu2011a,Yoshida2011,Yi2012,Ye2014} in which the hole pockets disappear and in hole doped systems \cite{Xu2013} in which the electron pockets disappear. In addition, ARPES has detected non-Fermi-liquid (linear in energy) scattering rates $\Gamma$ of the charge carriers independent of the control parameter\,\cite{Brouet2011,Rienks2013,Fink2015}. ARPES  is especially valuable since it can detect the momentum dependence and the orbital dependence of the scattering rates in multi band systems.  The scattering rates  and the related imaginary part of the self-energy $\Im\Sigma$\,\cite{Mahan2000} characterize the many-body properties of quasiparticles in solids.

In this article we first describe the appearance of a  linear-in-energy dependence of  $\Gamma$ in  systems close to a pocket vanishing  Lifshitz transition. Next we discuss the linear in energy scattering rates in correlated materials. Then we describe the fact that the non-Fermi-liquid behaviour near a Lifshitz transition is enhanced by correlation effects. On the basis of this understanding and the experimental ARPES results we conclude that a co-action of the non-Fermi liquid scattering rates with a Lifshitz transition of the pocket vanishing type, i.e.,  a crossing of a flat band through the Fermi level, leads to a huge mass enhancement. This scenario provides an alternative model to understand the strange normal state properties and possibly also unconventional superconductivity in correlated matter. 
\section{Scattering rates near a Lifshitz transition}
In this section  the scattering rate of the charge carriers $\Gamma_{ee}$ due to electro- electron interaction is described in a minimum model. In a first approximation the scattering rate is determined by an Auger process in the conduction band, i.e.,  a relaxation of an excited photo hole to lower energies relative to the Fermi level. The received energy is transfered to  
\begin{figure}[tb]
\onefigure [angle=0,width=\linewidth]{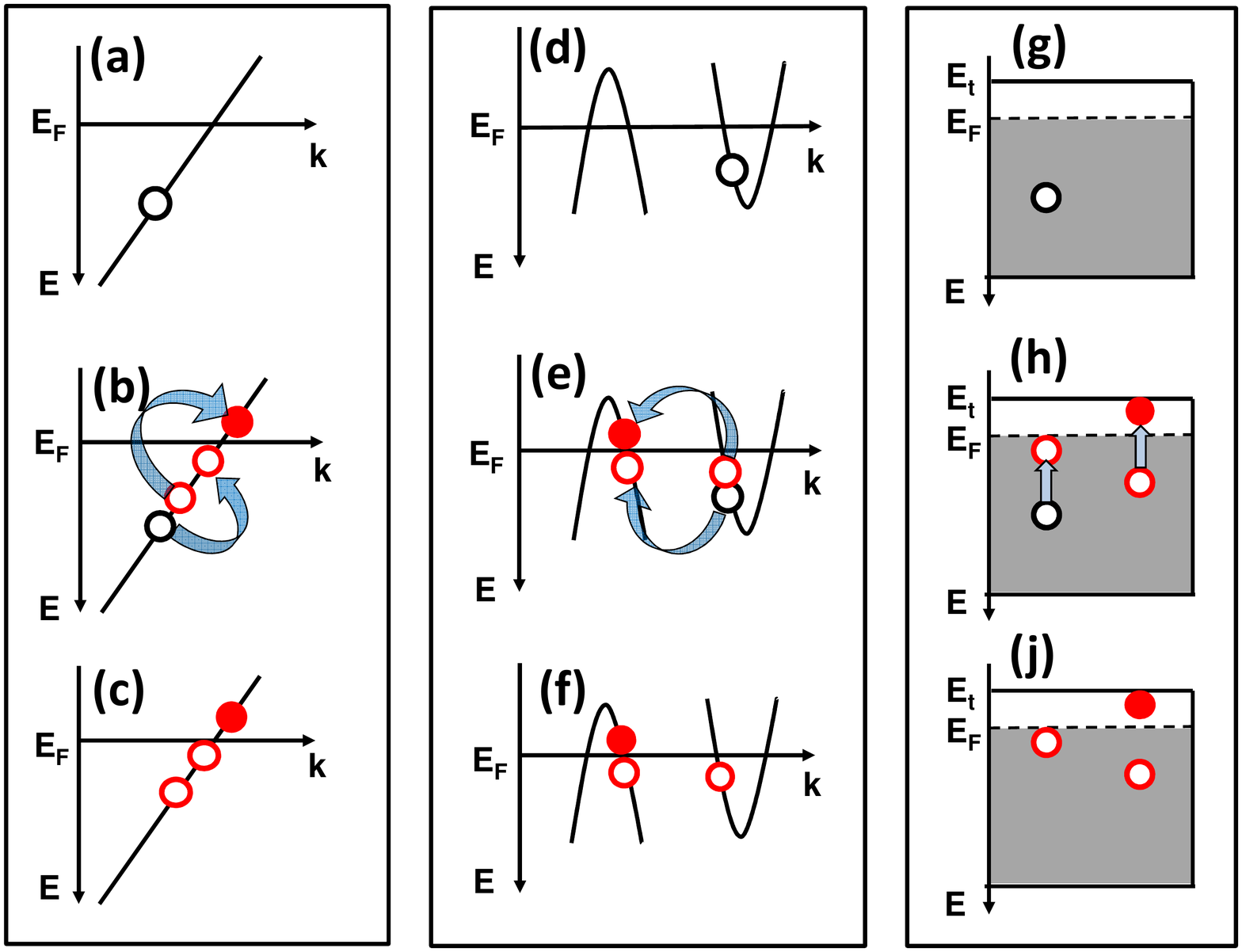}
\caption{
 (Color online) Electron-electron Auger-like scattering process. Top row: initial state with a photo hole. Middle row: Auger-like transitions. Lower row: final state with a relaxed photo hole and an electron-hole excitation.  (a-c) For a single conduction  band. (d-f) Same as (a-c) but for a band structure with hole and electron pockets to illustrate electron-electron scattering in ferropnictides. (g-j) Same as (a-c) but in a different presentation and for  a partially filled conduction band close to a Lifshitz transition. In this case the top of the conduction band at $E_t$ is close to the Fermi level.}
\label{fig1}
\end{figure}
 an electron-hole excitation (see fig~\ref{fig1}\,(a-c))\,\cite{Mahan2000}. Following this description the scattering rate can be calculated by the relation\,\cite{Mahan2000,Monney2012} 
\begin{equation}
\label{eq:gamma}
\begin{split}
\Gamma_{ee}(E,T)\propto U^2\sum_{\boldsymbol{Q}} \int \limits_{0}^{E} dE' \\
\times \xi(E',\epsilon_{\boldsymbol{k}},\epsilon_{\boldsymbol{k}+\boldsymbol{Q}},T)
 \chi(E',\epsilon_{\boldsymbol{k}},\epsilon_{\boldsymbol{k}-\boldsymbol{Q}},T).
 \end{split}
\end{equation}
$\boldsymbol{Q}$ is the momentum transfer for the excitations, $T$ is the temperature, $E$ the energy relative to the Fermi energy, and $\epsilon_{\boldsymbol{k}}$ is the bare particle dispersion. For the Coulomb interaction between two holes, a local approximation is used, i.e.,  it is described by the on-site Coulomb potential $U$. $\xi(E',\epsilon_{\boldsymbol{k}},\epsilon_{\boldsymbol{k}+\boldsymbol{Q}},T)$ describes the probability for the hole relaxation in the occupied part of the band while 
$\chi(E',\epsilon_{\boldsymbol{k}},\epsilon_{\boldsymbol{k}-\boldsymbol{Q}},T)$ describes the susceptibility for electron-hole excitations between the occupied and the unoccupied part of the band (e.g. the Lindhardt susceptibility\,\cite{Mahan2000}).
In the following, we do not evaluate the susceptibilities by taking into account the full band structure. To simplify matters, a  one band model is used (see fig~\ref{fig1}\,(a)). For the ferropnictides a two band model ((see fig~\ref{fig1}\,(d-f)) would be a better approximation. The sum over the momentum is neglected and $\xi(E',\epsilon_{\boldsymbol{k}},\epsilon_{\boldsymbol{k}+\boldsymbol{Q}},T)$ is approximated by 
 a constant density of states $D(0)$ at the Fermi level multiplied by the Fermi function $N_F(E,T)=1/(1+exp(E/k_BT))$ ($k_B$ is the Boltzman constant):
\begin{equation}
\label{eq:xi}
\xi \propto D(0) N_F(E,T).
\end{equation}

  $\chi(E,\epsilon_{\boldsymbol{k}},\epsilon_{\boldsymbol{k}-\boldsymbol{Q}},T)$  is approximated by  the convolution of the occupied part of the band with the unoccupied part of the band,  the Fermi edges of which being  broadened  by $N_F(E,T)$ and $1-N_F(E,T)$, respectively:
  \begin{equation}
\label{eq:chi}
\begin{split}
\chi(E,T)\propto \int dE\rq{}\\
\times D(E\rq{}) N_F(E\rq{},T) D(E-E\rq{}) (1-N_F(E-E\rq{},T)
 \end{split}
\end{equation}
For the occupied part of the band   $D(E\rq{})=D(0)$ is used while for the unoccupied part $D(E\rq{})=D(0)$ for $E\rq{}>E_t$  and $D(E\rq{})=0$ for $E\rq{}<E_t$  is used ($E_t$ is negative, see fig.~\ref{fig1}\,(g-i)).
By moving $E_t$ to zero we can simulate a  pocket vanishing type Lifshitz transition.

First we discuss the scattering rate for zero temperature and a wide conduction band ($E_t\ll 0$). In this case $\xi$ is constant up to the Fermi energy and the convolution of the two Fermi edges leads to $\chi$ which  is linear in energy. Performing the integration leads to quadratic increase of the scattering rate as a function of energy. This is observed in ARPES  of normal metals, such as Mo\,\cite{Valla1999} where the scattering rate for electron-electron interaction $\Gamma_{ee}$ is proportional to  $E^2$. This also leads to a quadratic temperature dependence of the resistivity in normal metals at low temperatures where phonons can no more be excited.
 
For the case near a Lifshitz transition ($E_t \to 0$) the unoccupied part is a
$\delta$ function at the Fermi level and therefore $\chi$ is constant in energy. Performing the integration yields a non-Fermi-liquid linear-in-energy scattering rate. 
\begin{figure}[tb]
\centering
\vspace{-0.5 cm}
\onefigure[width=\linewidth]{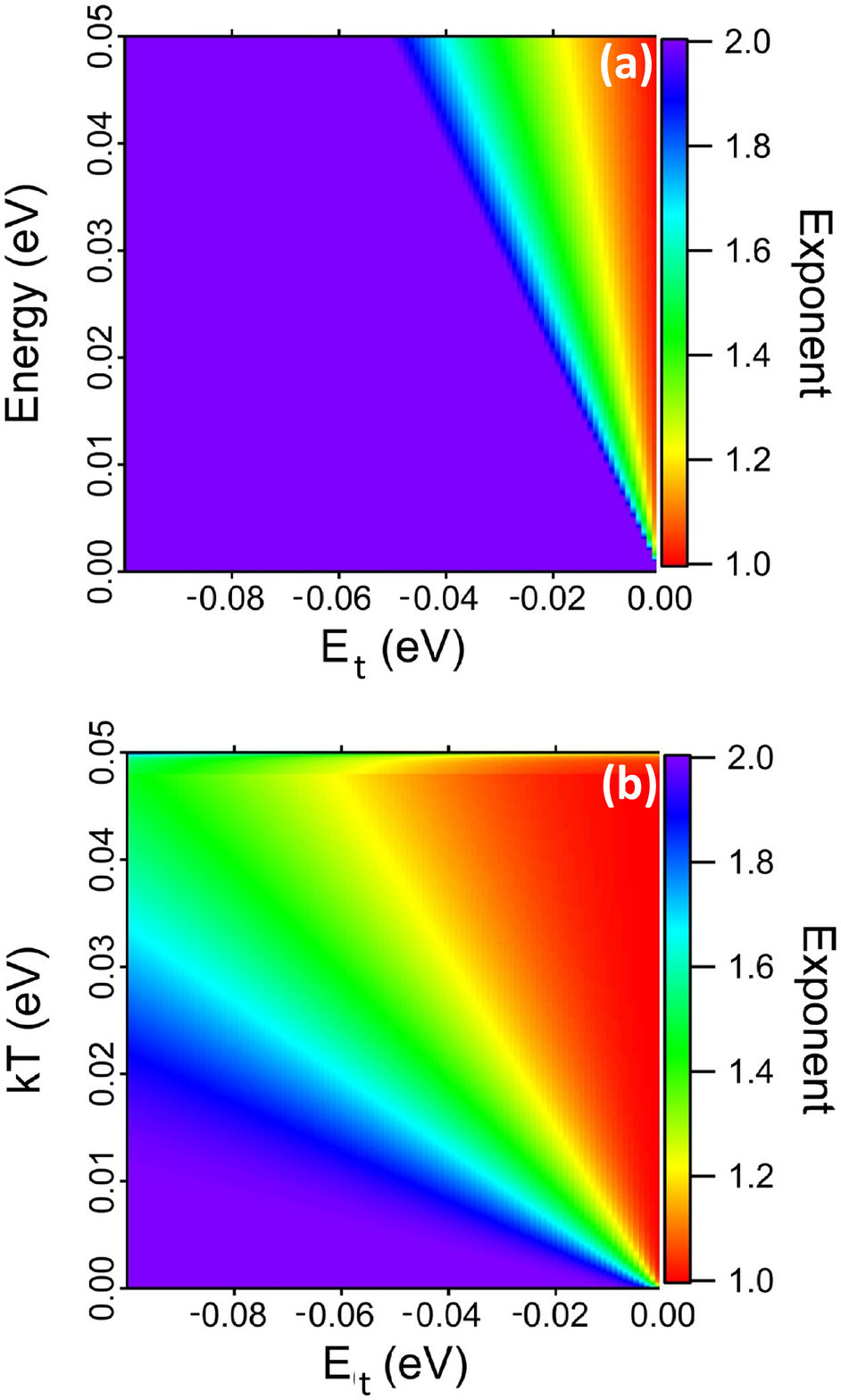}
\vspace{-0.5 cm}
\caption{ 
 (Color online) (a) False colour presentation of the exponent of the energy dependence of the scattering rate as a function of the top of the conduction band ($E_t)$ and the energy.
 (b) The exponent of the energy dependence  of the scattering rate as a function of the top of the conduction band ($E_t)$ and the thermal energy $k_BT$.} 
\label{fig2}
\end{figure}

In fig.~\ref{fig2}\,(a) we show the exponent of the energy dependence of the scattering rate as a function of $E_t$ and  the  binding energy. Near the Lifshitz transition ($E_t=0$) a non-Fermi liquid behaviour is obtained for all energies. With increasing distance to the Lifshitz transition (more negative $E_t$ values) the scattering rate transforms to a Fermi liquid behaviour with a quadratic energy dependence. A similar calculation was performed for the temperature dependence instead of the energy dependence (see fig.~\ref{fig2}\,(b)). Near the Lifshitz transition a linear-in-temperature non-Fermi-liquid behaviour is observed. With increasing distance to the Lifshitz transition a quadratic temperature Fermi-liquid behaviour is prevailing at low temperatures. We emphasize that in these calculations for an uncorrelated system, non-Fermi-liquid behaviour occurs only very close to the Lifshitz transition for $E>|E_t|$.
\section{Scattering rates in correlated systems}
\begin{figure} [tb]
\centering
\onefigure[angle=0,width=\linewidth]{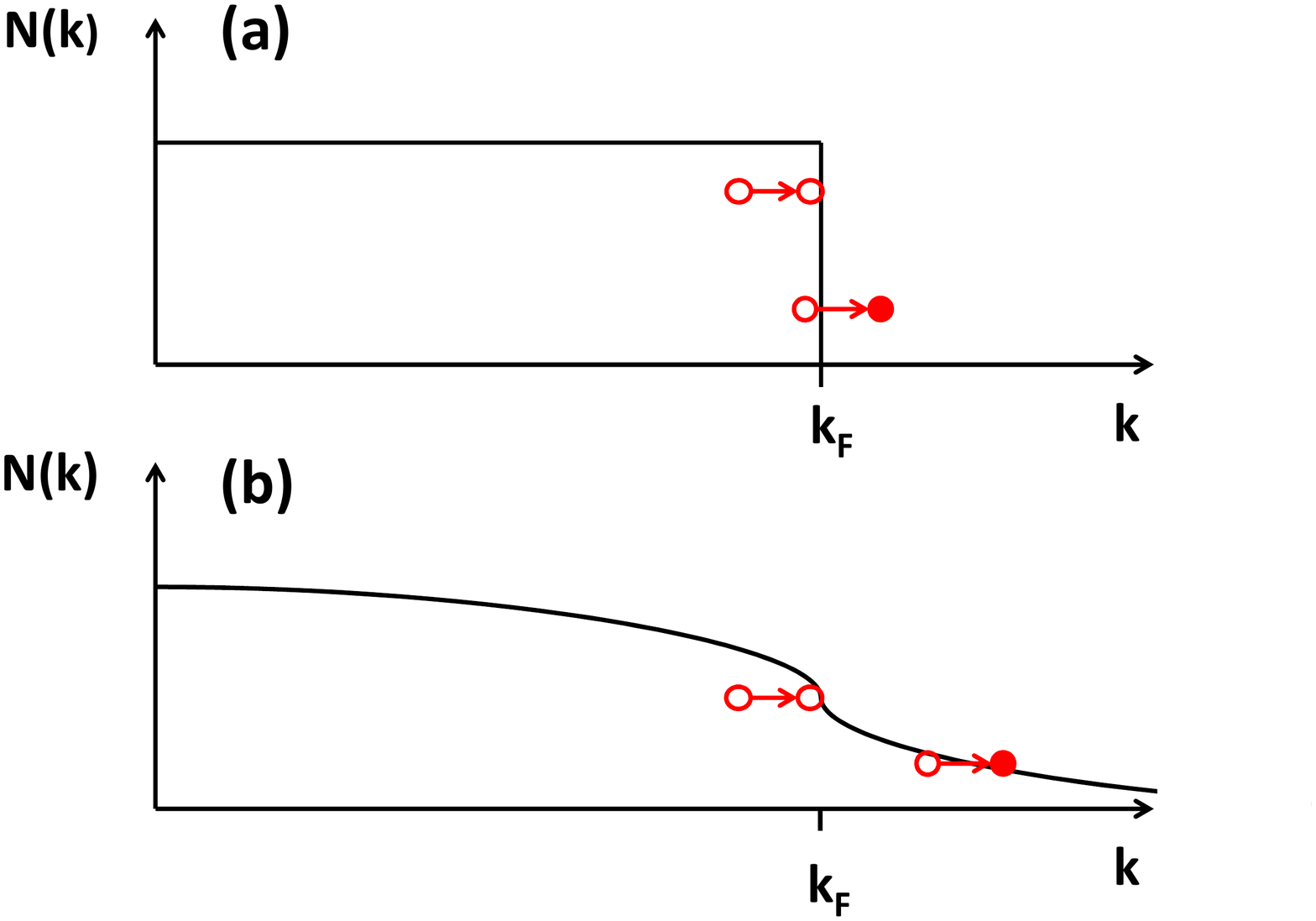}
\caption{ 
 (Color online) Momentum density $N(k)$  together with an Auger-like scattering process in the conduction band. (a) For a weakly correlated Fermi liquid. (b) For a  highly correlated metal.} 
\label{fig3}
\end{figure}
In this section we try to understand the difference of the energy dependence of the scattering rate between a normal and a highly correlated metal. In  fig.~\ref{fig3}\,(a) the momentum density of a weakly correlated  Fermi liquid together with the relaxation of an added photo hole due to an electron-hole excitation, already discussed in the previous section, is presented. The phase space for the hole  relaxation is proportional to the energy $E$. Because of the Pauli principle electrons can only excited with an energy E below the Fermi level. Thus the total phase space for the scattering process is proportional to $E^2$\,\cite{Mahan2000}, already demonstrated in the previous section.

In the case of strong correlation effects the hight of the step of the momentum  density at $k_F$ which is proportional to the renormalization factor $Z$ is reduced and large fractions are shifted from regions below $k_F$ to regions above $k_F$ (see fig.\,\ref{fig3}\,(b))\,\cite{Mahan2000}. In the case of very strong correlation effects, $Z$ which determines the weight of coherent quasiparticles, is small and a large fraction (proportional to $1-Z$) is due to  incoherent particles. When we add a coherent quasiparticle with energy $E$ to the system (e.g. a photo hole)   the phase space for the decay of the quasiparticle is again proportional to $E$. However, in this case the relaxation energy can be transferred also to the incoherent charge carriers.  These can be excited not only within a narrow energy range  $E$ below the Fermi level but the electron-hole excitations of the incoherent charge carriers,  which are no more restricted by the Pauli principle, can now occur in the entire range of the band width $W$. Thus in this case the total phase space for the relaxation process is proportional to $EW$ which for $E \ll W$ is proportional to $E$. This leads to the marginal Fermi liquid model in which one assumes that the weight of the coherent particles is close to zero, i.e., $Z \ll 1$. Thus different from a normal Fermi liquid, the interaction of the coherent quasiparticles with the incoherent particles yield  strong low-energy relaxation processes. In this view it is difficult to rationalize a Fermi liquid behaviour of a strongly correlated single band metal since the total relaxation process is a sum of two components, one which proportional to $ZE^2$ and a second one which is proportional $(1-Z)E$.
\section{Scattering rates of a correlated system near a Lifshitz transition}
In this Section  the co-action of a Lifshitz transition and correlation effects is discussed using the ARPES results of the electronic structure of the hole pocket in ferropnictides\,\cite{Fink2015}.
\begin{figure}[tb]
\onefigure[width=\linewidth]{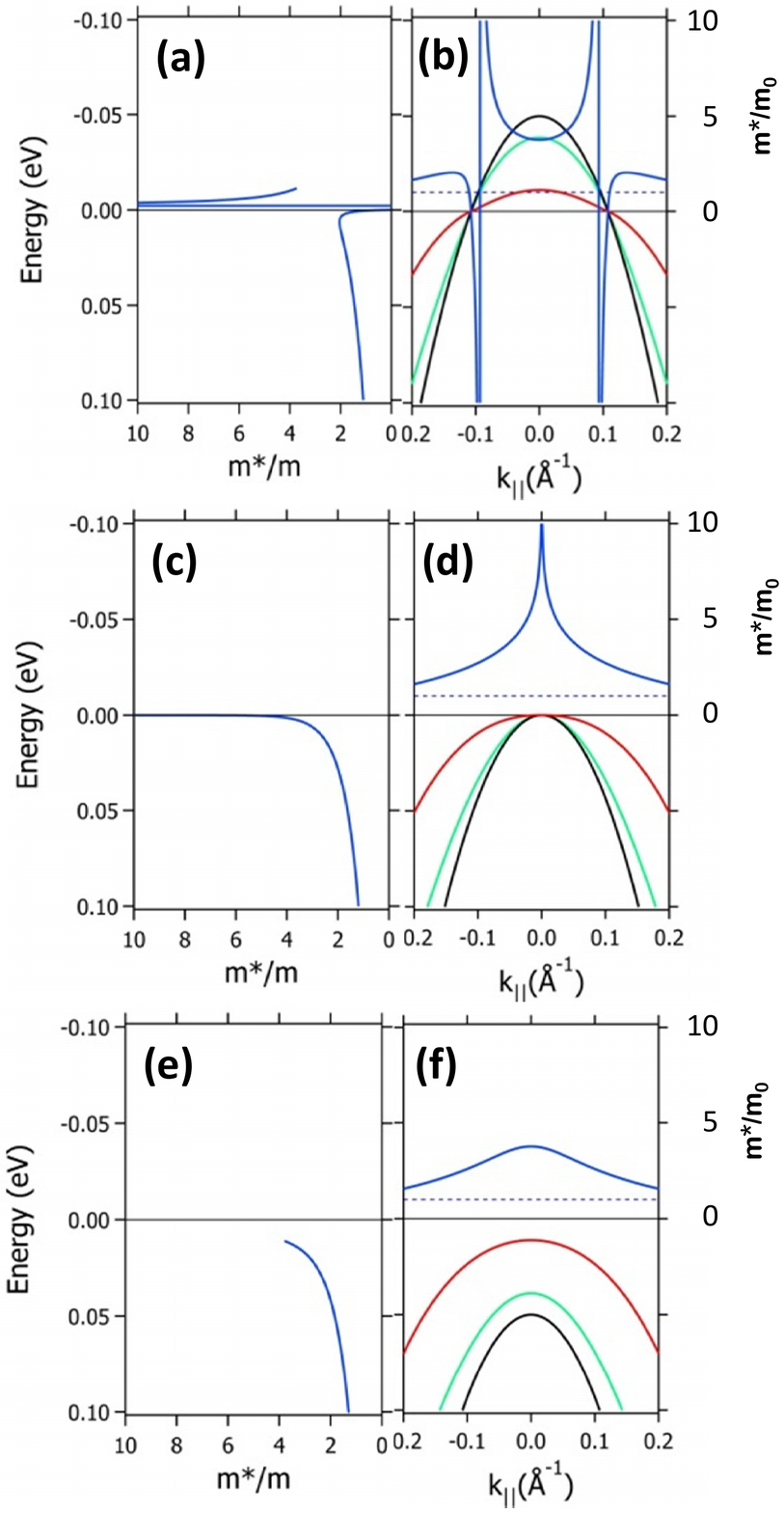}
\caption{ 
 (Color online) Calculation of the renormalized dispersion (red line) and the effective mass $m^*/m_0$  (blue lines)  as a function of energy (left panels) and as a function of momentum (right panels) using a bare particle parabolic dispersion (black line)  and a $\Re\Sigma$ derived from ARPES experiments on ferropnictides. (a) and (b) For $E_t =  -50$ meV, (c) and (d) For $E_t =  0$, (e) and (f) For $E_t =  50$  meV.} 
\label{fig4}
\end{figure}
In these experiments on  electron doped compounds, a linear-in-energy  dependence of $\Gamma$ was observed in a large range of the control parameter. It is assume that the experimental results for $\Gamma$ can be extrapolated to low energies (well below a typical ARPES energy resolution of about 5 meV). From $\Gamma$ the imaginary part of the self-energy $\Im\Sigma$ is calculated by the relation $\Gamma=Z\Im\Sigma$\,\cite{Mahan2000}. At high energies we assume that above a cutoff energy $E_c$, corresponding to the band width, $\Im\Sigma$ remains constant. Thus the complex self-energy  $\Sigma$ corresponds to that proposed in the marginal Fermi liquid model\,\cite{Varma2002}:
\begin{equation}
\label{resi}
\Sigma_{MF}(E)=\frac{1}{2}\lambda_{MF}E ln\frac{E_c}{x}+i \frac{\pi}{2}\lambda_{MF}x.
\end{equation}
$x=max(|E|,k_BT)$ and  $k_BT$ is the thermal energy. $\lambda_{MF}$ is a dimensionless coupling constant.  In the calculations presented in fig.\,\ref{fig4}  zero thermal energy and typical values derived from ARPES experiments on the inner hole pocket of electron doped ferropnictides\,\cite{Fink2015} are used: $\lambda_{MF}= 1.4$ and $E_c=1.5$ eV. The parabolic  bare particle dispersion (black lines in  fig.\,\ref{fig4}) was taken from a fit to bands of the inner hole pocket of ferropnictides derived from DFT band structure calculations. Using the above given parameters, $\Re\Sigma$ was calculated with the help of eq.~(\ref{resi})  (green  lines in  fig.\,\ref{fig4}). The difference between the bare particle dispersion and $\Re\Sigma$  yields the renormalised dispersions\,\cite{Mahan2000} depicted in  fig.\,\ref{fig4} by red lines. In fig.\,\ref{fig4} we also present the effective masses $m^*/m_0$ (blue lines), derived from the second derivative of the renormalized dispersion. In fig.\,\ref{fig4} (b), (d), and (f)  $m^*/m_0$  is presented as a function of the momentum $k$. Similar data have been presented already  in Ref.\,\cite{Fink2015}. Here in fig.\,\ref{fig4} (a), (c), and (e) $m^*/m_0$ values as a function of energy are added.

In the case where the top of the bare particle hole pocket just touches the Fermi level, i.e., when at the Fermi level the bare particle dispersion is flat  (fig.\,\ref{fig4}\,(d)),  which occurs at the Lifshitz transition,  $\Re\Sigma$ is very close to the bare particle band and therefore  at the Fermi level the renormalized band is very flat. This corresponds to a very high effective mass $m^*/m_0 \approx 8$ at $k=0$
(fig.\,\ref{fig4}\,(d)) and at the Fermi level  (fig.\,\ref{fig4}\,(c)). Moving away from $k=0$ or $E=0$ the effective mass is strongly reduced to a value between 1 and 2.

When we shift the top of the hole pocket 50 meV above the Fermi level (fig.\,\ref{fig4}\,(a) and (b)), which in real systems corresponds to a changing of the control parameter by doping, by chemical pressure, or by pressure, the dispersion of the renormalized band is also reduced. However, the slope of this band at the Fermi level is strongly reduced when compared with
the case where the bare band touches the Fermi level. This reduction of the Fermi velocity is connected with the effective masses at $E_F$ which is reduced to a value of about 2.

Finally, when we shift the top of the hole pocket 50 meV below the Fermi level (fig.\,\ref{fig4}\,(e) and (f)), the band is still renormalized but the effective mass at $k=0$ is reduced to a value of about 4.

Summarizing the results of the calculations presented in fig.\,\ref{fig4},  at the Lifshitz transition and only at this transition one obtains  a correlation induced particularly enhanced flattening of the renormalized band which is related to an almost diverging effective mass at the Fermi level and at the momentum where the Lifshitz transition occurs. The correlation effects also lead to a pinning of the top of the hole pocket at the Fermi level.

\begin{figure}
\onefigure[width=\linewidth]{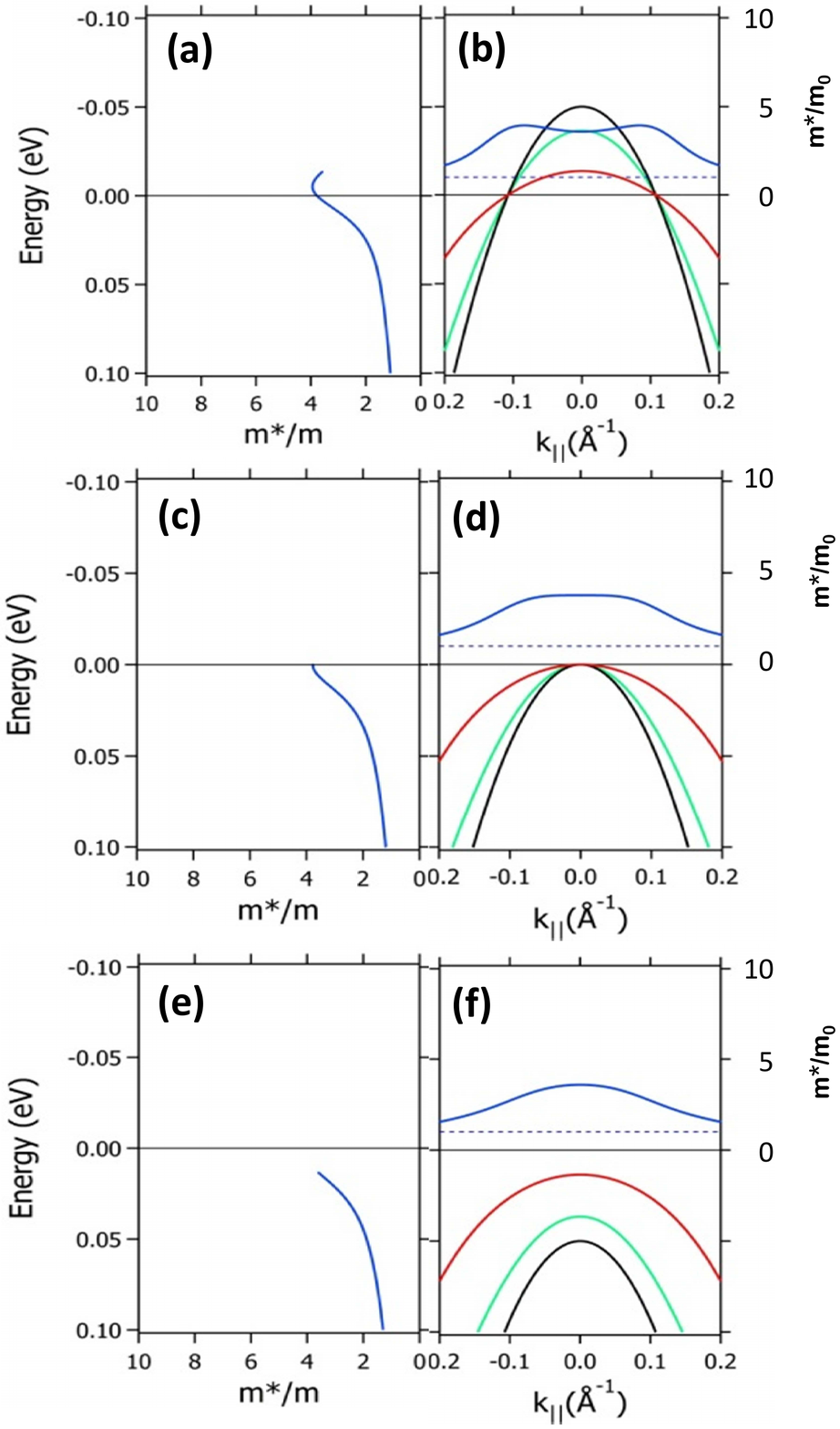}
\caption{ 
 (Color online) The same calculation as in fig.\,\ref{fig4}, but with the addition of a finite thermal energy of 0.03 eV.} 
\label{fig5}
\end{figure}

In fig.\,\ref{fig5} we present similar calculations but with a thermal energy of 0.03 eV corresponding to a temperature of 348 K. In the case of a Lifhitz transition (fig.\,\ref{fig5}\,(c) and (d)) the effective mass at the Fermi level is reduced form about 8 to 4. However away from the Lifshitz transition, when we shift the top of the hole pocket to -50 meV (fig.\,\ref{fig5}\,(a) and (b)) the effective mass at the Fermi level is enhanced from about 2 to 4 due to a broadening of the maximum above the Fermi level. 

\begin{figure}[tb]
\onefigure [width=0.8\linewidth]{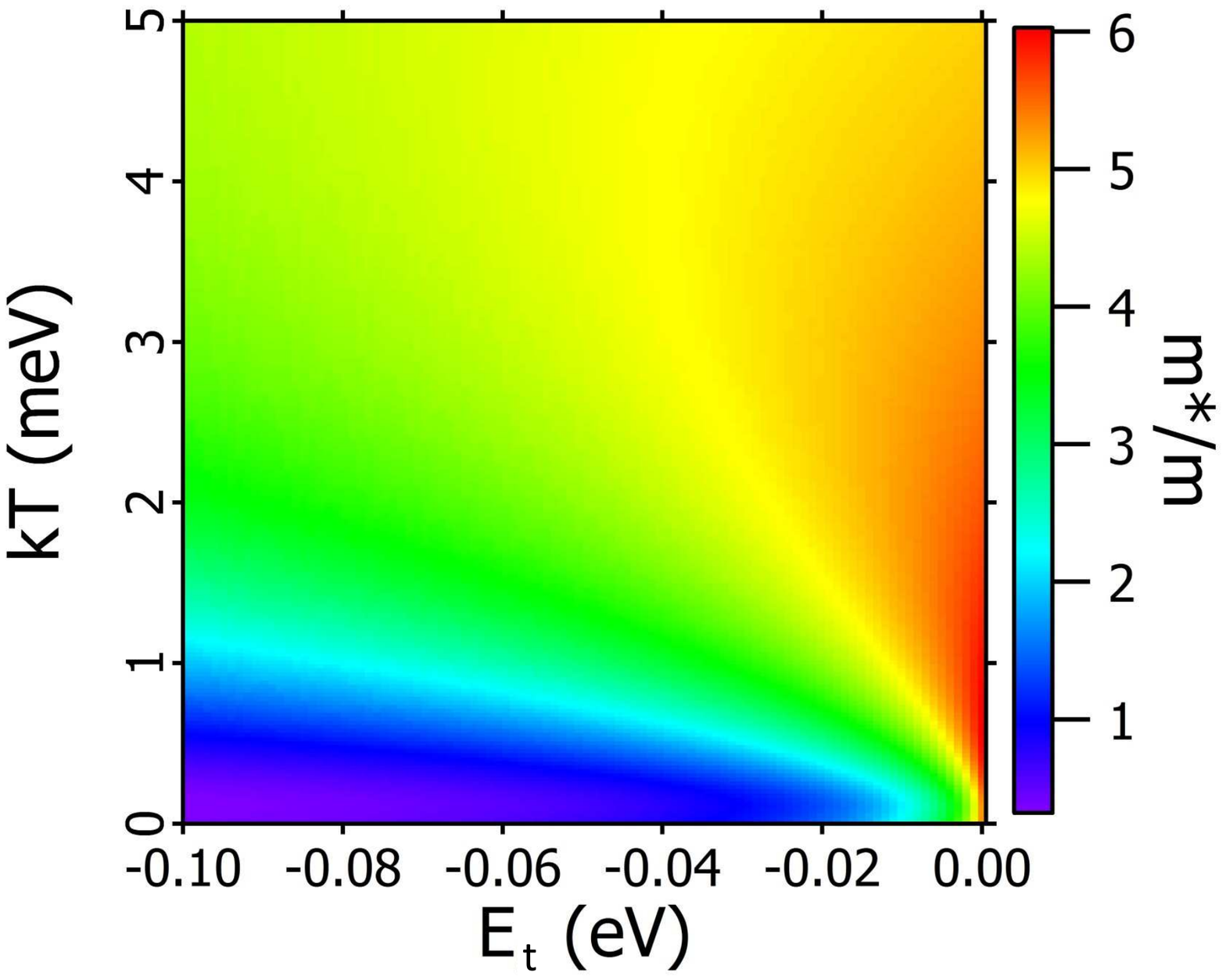}
\caption{ 
 (Color online) Calculation of the effective mass as a function of thermal energy $k_BT$ and a shift of the top of the hole band ($E_t$).} 
\label{fig6}
\end{figure}

Finally a calculation of the effective mass as a function of the thermal energy $k_BT$ and the shift of the top of the hole pocket is presented in fig.\,\ref{fig6}. An enhancement of the effective mass near the Lifshitz transition and at low temperatures is realized.

\section{Discussion }
 The present calculations offer a new explanation of   the non-Fermi-liquid behaviour of correlated systems  near a QCP which is different from the traditional view related to a coupling of the charge carriers to quantum spin fluctuations. In the present scenario the strange normal state properties follow from a co-action of a Lifshitz transition and strong local electron-electron interaction which lead to a large number if incoherent and a small number of coherent charge carriers.  At the end of the $3d$-row, correlation effects are caused by the Mott-Hubbard on-site Coulomb interaction while Hund's-rule  exchange interaction is more important in the middle of the row. That local on-site correlation effects are dominant in the scattering rates related to electron-electron interaction 
 is also supported by  theoretical calculations in the framework of DFT combined with dynamical mean-field theory (DMFT). Interestingly, non-Fermi-liquid-like self-energies have been reported\,\cite{Werner2012,Yin2012}  which could be fitted by sub-linear power laws over a range of elevated energies. The incoherent-metal behavior has been attributed to an interplay of Hubbard and Hund's-rule couplings in a multi-band system.

In the following  we compare the present calculations with normal state transport and thermal properties of ferropnictides. 
Several investigations on \BFAP\  by  London penetration depth\,\cite{Hashimoto2012,Walmsley2013} and  de Haas-van Alphen effect\,\cite{Shishido2010} measurements have detected a huge mass enhancement at the QCP which was interpreted by a coupling to antiferromagnetic quantum fluctuations. The present calculations provide an alternative explanation of the mass enhancement in terms of a co-action of correlation effects and a Lifshitz transition (see fig.~\ref{fig6}).  Close to the QCP the resistivity data on \BFAP\  \,\cite{Kasahara2010,Analytis2014} show  a linear temperature dependence over a large range, but moving away from the QCP a Fermi liquid behaviour is detected at low temperatures. It is difficult to interpret these results on the basis of the present calculations since the conductivity depends in a non-trivial way not only on the  scattering rate but also on the effective mass and the number of charge carriers.  The present  calculations reveal that all three parameter are changing near a Lifshitz transition. According to previous work\,\cite{Grimvall1981} the conductivity should be not renormalized  by many-body interactions. Possibly the ARPES results could be related to  \lq\lq{}hot\rq\rq{} sections of the Fermi surface mediating antiferromagnetism and superconductivity while \lq\lq{}cold\rq\rq{} sections, not detected in the ARPES experiments, could be related to the normal state conductivity.
Possibly also the Fermi-liquid behaviour of the ferropnictides detected in the optical conductivity\,\cite{Tytarenko2015} could be related to the \lq\lq{}cold\rq\rq{} spots . In this context one should mention that the conductivity of electron doped ferropnictides is predominately determined by electronic states of the electron pockets\,\cite{Rullier-Albenque2016}, which may be less correlated because there is no enhancement by the proximity to a Lifshitz transition.  The thermoelectric power is connected to the derivative of the density of electronic states at the Fermi level. Thus the divergent thermopower detected in \BFCA\ near optimal doping\,\cite{Arsenijevic2013} can be readily explained by the scenario described in this contribution: a coaction of a Lifshitz transition and correlation effects. The present calculations of the mass enhancement at the Lifshitz transition  can probably also explain the thermal expansion data and the associated Gr\"uneisen parameter of \BFCA\ \,\cite{Meingast2012}. Furthermore, also the thermal data on the superconducting heat capacity jump in \BFAP\ \,\cite{Analytis2014} can be explained within this model. We also mention that  the enhanced relaxation rate $1/T_1T$ near the QCP detected in NMR experiments on \BFCA\ can be  related to a Lifshitz transition\,\cite{Ning2010}.  Moreover, the data  for the effective mass as a function of the control parameter, i.e., the band shift, resemble very much data on the temperature dependence of the resistivity as  a function of the control parameter in heavy Fermion systems\,\cite{Gegenwart2008} or ferropnictides\,\cite{Analytis2014}. 

As already mentioned previously\,\cite{Fink2015}, the high effective mass derived at the Lifshitz transition is related to  an effective small Fermi energy. This can lead to a break-down of the Migdal theorem\,\cite{Migdal1958} and an interpolating superconducting state between BCS theory and BE condensation\,\cite{Bose1924}.
\section{Summary }
We have performed calculations of the effective mass for a highly correlated metal with a non-Fermi-liquid behaviour as a function of the the distance of a flat band relative to the Fermi level. From these calculations we derive a model in which the strange normal state transport and thermal properties near optimal substitution/doping can be explained by  a co-action of a marginal Fermi liquid self-energy with a van Hove singularity  at the Fermi level which is expected near a hole vanishing Lifshitz transition.  The high effective masses imply small effective  Fermi energies which could be comparable to bosonic energies mediating superconducting pairing. Thus possible the Migdal's theorem is violated in the  unconventional superconductors which could lead to a phase which is  near the BCS-BE crossover. The results can be generalised to other unconventional superconductors and possibly are a recipe  for future search of new high-$T_c$ superconductors.

\acknowledgments
 This work was financially supported by the German Research Foundation the DFG through the priority program SPP1458. Valuable discussions with S. Borisenko, B. B\"uchner, P. Fulde,  D. van der Marel, and C. Monney are acknowledged.


\begin{thebibliography}{10}
\bibliographystyle{phaip}

\bibitem{Loehneysen2007}
H.~v. L\"ohneysen, A.~Rosch, M.~Vojta, and P.~W\"olfle,
\newblock Rev. Mod. Phys. {\bf 79}, 1015 (2007).

\bibitem{Gegenwart2008}
P.~Gegenwart, Q.~Si, and F.~Steglich,
\newblock Nat Phys {\bf 4}, 186 (2008).

\bibitem{Haslinger2002}
R.~Haslinger, A.~Abanov, and A.~Chubukov,
\newblock Europhys. Lett. {\bf 58}, 271 (2002).

\bibitem{Custers2003}
J.~Custers et~al.,
\newblock Nature {\bf 424}, 524 (2003).

\bibitem{Daou2009}
R.~Daou et~al.,
\newblock Nat Phys {\bf 5}, 31 (2009).

\bibitem{Kasahara2010a}
S.~Kasahara et~al.,
\newblock Phys. Rev. B {\bf 81}, 184519 (2010).

\bibitem{Analytis2014}
J.~G. Analytis et~al.,
\newblock Nat Phys {\bf 10}, 194 (2014).

\bibitem{Varma2002}
C.~Varma, Z.~Nussinov, and W.~van Saarloos,
\newblock Physics Reports {\bf 361}, 267 (2002).

\bibitem{Damascelli2003}
A.~Damascelli, Z.~Hussain, and Z.-X. Shen,
\newblock Rev. Mod. Phys. {\bf 75}, 473 (2003).

\bibitem{Johnston2010}
D.~C. Johnston,
\newblock Adv. Phys. {\bf 59}, 803 (2010).

\bibitem{Lifshitz1960}
I.~M. Lifshitz,
\newblock Sov. Phys. JETP {\bf 11}, 1130 (1960).

\bibitem{Malaeb2009}
W.~Malaeb et~al.,
\newblock J. Phys. Soc. Jpn. {\bf 78}, 123706 (2009).

\bibitem{He2010}
C.~He et~al.,
\newblock Phys. Rev. Lett. {\bf 105}, 117002 (2010).

\bibitem{Thirupathaiah2010}
S.~Thirupathaiah et~al.,
\newblock Phys. Rev. B {\bf 81}, 104512 (2010).

\bibitem{Thirupathaiah2011}
S.~Thirupathaiah et~al.,
\newblock Phys. Rev. B {\bf 84}, 014531 (2011).

\bibitem{Liu2011a}
Z.-H. Liu et~al.,
\newblock Phys. Rev. B {\bf 84}, 064519 (2011).

\bibitem{Yoshida2011}
T.~Yoshida et~al.,
\newblock Phys. Rev. Lett. {\bf 106}, 117001 (2011).

\bibitem{Yi2012}
M.~Yi et~al.,
\newblock New Journal of Physics {\bf 14}, 073019 (2012).

\bibitem{Ye2014}
Z.~Ye et~al.,
\newblock Phys. Rev. X {\bf 4}, 031041 (2014).

\bibitem{Xu2013}
N.~Xu et~al.,
\newblock Phys. Rev. X {\bf 3}, 011006 (2013).

\bibitem{Brouet2011}
V.~{Brouet} et~al.,
\newblock arXiv e-prints:1105.5604  (2011).

\bibitem{Rienks2013}
E.~D.~L. Rienks et~al.,
\newblock EPL (Europhysics Letters) {\bf 103}, 47004 (2013).

\bibitem{Fink2015}
J.~Fink et~al.,
\newblock Phys. Rev. B {\bf 92}, 201106 (2015).

\bibitem{Mahan2000}
G.~D. Mahan,
\newblock {\em Many-Particle Physics},
\newblock Kluwer Academic/Plenum Publishers, New York, 2000.

\bibitem{Monney2012}
C.~Monney, G.~Monney, P.~Aebi, and H.~Beck,
\newblock Phys. Rev. B {\bf 85}, 235150 (2012).

\bibitem{Valla1999}
T.~Valla, A.~V. Fedorov, P.~D. Johnson, and S.~L. Hulbert,
\newblock Phys. Rev. Lett. {\bf 83}, 2085 (1999).

\bibitem{Werner2012}
P.~Werner et~al.,
\newblock Nat Phys {\bf 8}, 331 (2012).

\bibitem{Yin2012}
Z.~P. Yin, K.~Haule, and G.~Kotliar,
\newblock Phys. Rev. B {\bf 86}, 195141 (2012).

\bibitem{Hashimoto2012}
K.~Hashimoto et~al.,
\newblock Science {\bf 336}, 1554 (2012).

\bibitem{Walmsley2013}
P.~Walmsley et~al.,
\newblock Phys. Rev. Lett. {\bf 110}, 257002 (2013).

\bibitem{Shishido2010}
H.~Shishido et~al.,
\newblock Phys. Rev. Lett. {\bf 104}, 057008 (2010).

\bibitem{Kasahara2010}
S.~Kasahara et~al.,
\newblock Phys. Rev. B {\bf 81}, 184519 (2010).

\bibitem{Grimvall1981}
G.~Grimvall,
\newblock {\em The electron-phonon interaction in metals},
\newblock North-Holland, 1981.

\bibitem{Tytarenko2015}
A.~Tytarenko, Y.~Huang, A.~de~Visser, S.~Johnston, and E.~van Heumen,
\newblock Scientific Reports {\bf 5}, 12421 (2015).

\bibitem{Rullier-Albenque2016}
F.~Rullier-Albenque,
\newblock Comptes Rendus Physique {\bf 17}, 164 (2016).

\bibitem{Arsenijevic2013}
S.~Arsenijevic et~al.,
\newblock Phys. Rev. B {\bf 87}, 224508 (2013).

\bibitem{Meingast2012}
C.~Meingast et~al.,
\newblock Phys. Rev. Lett. {\bf 108}, 177004 (2012).

\bibitem{Ning2010}
F.~L. Ning et~al.,
\newblock Phys. Rev. Lett. {\bf 104}, 037001 (2010).

\bibitem{Migdal1958}
A.~B. Migdal,
\newblock Soviet Physics $\mathrm{JETP-USSR}$ {\bf 34}, 996 (1958).

\bibitem{Bose1924}
S.~N. Bose,
\newblock Z. phys {\bf 26}, 178 (1924).



\end{thebibliography}
\end{document}